\def\DI{{D_{\text{I}}}}
\def\Dh{{D_{\text{H}}}}

\documentclass[a4paper,fleqn]{cas-dc}

\usepackage[numbers]{natbib}

\def\tsc#1{\csdef{#1}{\textsc{\lowercase{#1}}\xspace}}
\tsc{WGM}
\tsc{QE}
\tsc{EP}
\tsc{PMS}
\tsc{BEC}
\tsc{DE}

\begin{document}
\let\WriteBookmarks\relax
\def\floatpagepagefraction{1}
\def\textpagefraction{.001}
\shorttitle{}
\shortauthors{}

\title [mode = title]{Thermal noise effects on the magnetization switching of a ferromagnetic anomalous Josephson junction}

\author[1]{C Guarcello}[orcid=0000-0002-3683-2509]
\cormark[1]
\ead{cguarcello@unisa.it}

\address[1]{Dipartimento di Fisica ``E.R. Caianiello'', Universit\`a di Salerno, Via Giovanni Paolo II, 132, I-84084 Fisciano (SA), Italy}

\author[2,3]{FS Bergeret}[orcid=0000-0001-6007-4878]
\ead{sebastian_bergeret@ehu.eus}

\address[2]{Centro de Física de Materiales, Centro Mixto CSIC-UPV/EHU, Paseo Manuel de Lardizabal 5, 20018 San Sebastián, Spain}

\address[3]{Donostia International Physics Center, Paseo Manuel de Lardizabal 4, 20018 San Sebastián, Spain}

\cortext[cor1]{Corresponding author}

\begin{abstract}
We discuss the effects of thermal noise on the magnetic response of a lateral ferromagnetic Josephson junction with spin-orbit coupling and out-of-plane magnetization. 
The direction of the magnetic moment in the ferromagnetic layer can be inverted by using controlled current pulses. This phenomenon is due to the magnetoelectric effect that couples the flowing charge current and the magnetization of the ferromagnet. We investigate the magnetization reversal effect versus intrinsic parameters of the ferromagnet, such as the Gilbert damping and strength of the spin-orbit coupling. We estimate the magnetization reversing time and find the optimal values of the parameters for fast switching.
With the aim of increasing the operation temperature we study the effects induced by thermal fluctuations on the averaged stationary magnetization, and find the conditions that make the system more robust against noise.
\end{abstract}

\begin{keywords}
Ferromagnetic Josephson junction \sep Anomalous Josephson effect \sep Magnetization reversal phenomenon \sep Landau–Lifshitz–Gilbert (LLG) equation \sep Resistively shunted junction (RSJ) model \sep Thermal noise effects
\end{keywords}

\maketitle

\section{Introduction}
\label{Sec01}

In the past few years many efforts have been devoted to the theoretical study of the magnetic response of ferromagnetic anomalous Josephson
junctions (JJs)~\cite{Kul14,Shu17,Nas18,ShuMaz18,Shu18,AtaPan19,Shu19,ShuNas19,Nas19,Ata19,NasShu19,Gua20,Rah20,Maz20}, thus offering a path of concrete applications based on the electrical control of the magnetization in a so-called $\varphi_0$--junction. A realization of such junction consists essentially of a superconductor-ferromagnet-superconductor (SFS) Josephson junction with an intrinsic spin-orbit coupling (SOC). Its ground state corresponds to a finite phase shift, $0<\varphi_0<\pi$, in the current-phase-relation. Recently, such anomalous phase has been observed experimentally in hybrid Josephson devices fabricated with a topological insulator Bi$_2$Se$_3$ and Al/InAs heterostructures and nanowires~\cite{Szo16,Ass19,May19,Str20}. As demonstrated theoretically~\cite{Buz08,Kon09,Shu17}, the magnetization of the F layer can be electrically controlled. In fact, in a $\varphi_0$--junction, due to the magnetoelectric effect, a charge current induces an in-plane magnetic moment~\cite{Ede95,Ede05,Mal08,Ber15,Kon15,Bob17}, which in turn acts as a torque on the out-of plane magnetization of the F layer, inducing eventually its switching. Alternatively, in the place of an electric current, the magnetization reversal can be driven by a magnetic field in a SQUID setup~\cite{Shu18}, \emph{i.e.}, a device widely used for detecting anomalous Josephson effects~\cite{Szo16,Ass19,May19,Str20,Mir20,GuaCit20}. 

The magnetization reversal phenomenon might eventually find an application in different fields of superconducting spintronics~\cite{Lin15,Esc15,Gol17}. Nevertheless, any concrete applications based on the magnetization of a $\varphi_0$ junction has to deal with the effects on the magnetic response stemming from the unavoidable thermal fluctuations. In fact, the temperature at which the system resides can significantly influence the evolution of the magnetic moment, even inducing unwanted transitions or hindering the inversion of the magnetization. 

The impact of stochastic thermal fluctuations on the magnetization reversal phenomenon was addressed in Ref.~\cite{Gua20}. That work describes also the feasibility, initially suggested in Ref.~\cite{Shu17}, to employ a current-biased SFS Josephson junction as a memory element, with the information encoded in the magnetization direction of the F layer. 

In the present work, we take a cue from Ref.~\cite{Gua20} and focus in more detail on the noise effects on the magnetization dynamics, studying how the value of some system parameters can influence the robustness against thermal fluctuations of the current-induced magnetization reversal phenomenon. 
In particular, we demonstrate that the values of the Gilbert damping parameter and strength of the spin-orbit coupling can be conveniently chosen to make the system more stable, to increase the range of suitable working temperatures.

The work is organized as follows: In Sec.~\ref{Sec02}, we present the theoretical model used to describe the time evolution of both the magnetic moment and the Josephson phase of a current-driven SFS junction. In Sec.~\ref{Sec03} we discuss the magnetic configuration of the junction after applying a current pulse. We look the stationary magnetization as a function of the Gilbert damping parameter and the SOC strength; additionally, we show a simple way to predict the overall response of the stationary magnetization. Moreover, we investigate the full temporal evolution of the magnetization and we demonstrate the possibility to minimize the magnetization switching time by adjusting specific system parameters. In Sec.~\ref{noiseeffects}, we show the effects of stochastic thermal fluctuations on the magnetization dynamics. Finally, in Sec.~\ref{Sec04} we present our conclusions. 

\begin{figure}[t!!]
\centering
\includegraphics[width=\columnwidth]{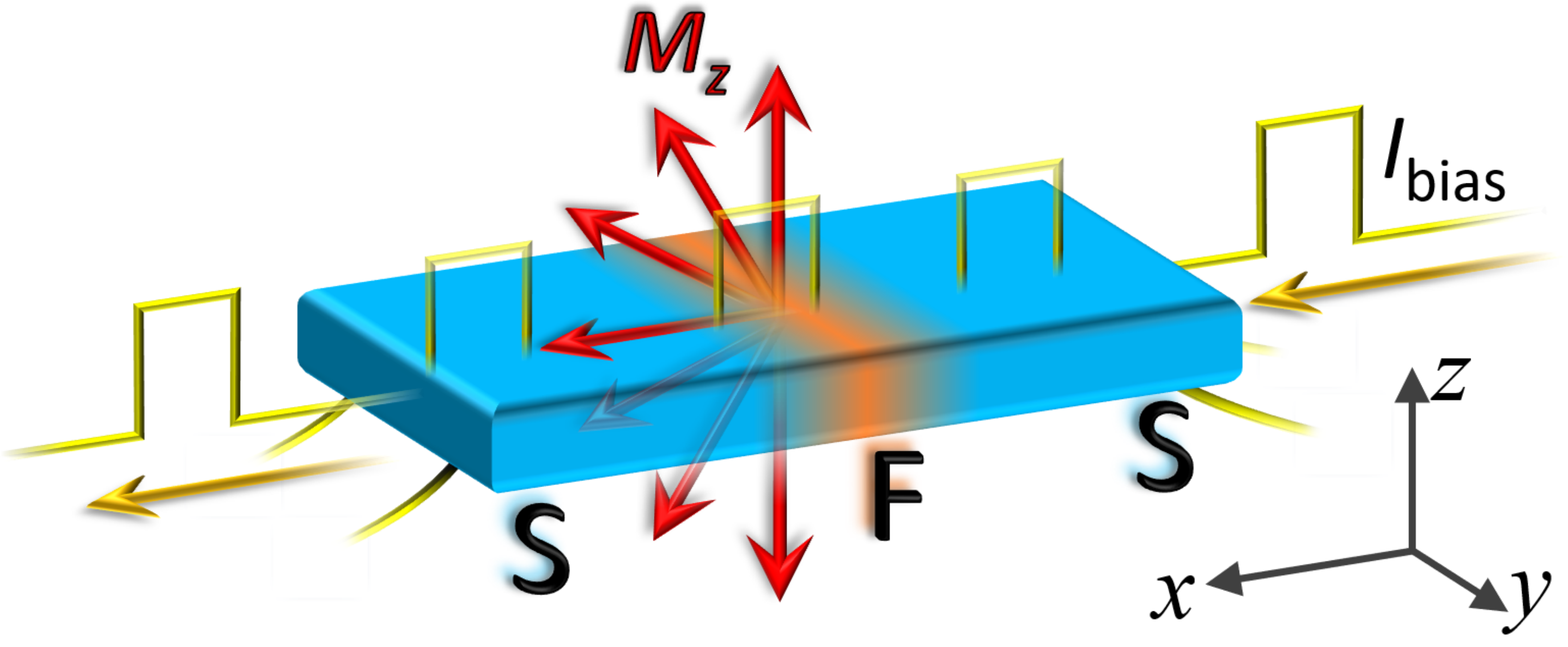}
\caption{S/F/S Josephson junction driven by a rectangular bias current pulse, $I_{\text{bias}}$. The cartoon depicts the inversion of the $z$-component of the magnetization of the F layer after the passage of the current pulse $I_{\text{bias}}$ through the junction.} 
\label{Fig00}
\end{figure}

\section{The model}
\label{Sec02}

The setup that we consider consists of an SFS junction, see Fig.~\ref{Fig00}, with a thin ferromagnetic film with an out-of-plane magnetic anisotropy and a Rashba-like SOC~\cite{Gua20}. 

Due to the interplay between the exchange field and the SOC, the current-phase relation of the SFS Josephson junction reads $I_{\varphi}=I_c\sin(\varphi-\varphi_0)$, where $I_c$ is the critical current of the junction, $\varphi$ is the Josephson phase difference, and $\varphi_0$ is the anomalous phase shift. The latter depends on several parameters of the system, such us the Rashba coefficient $\alpha$~\cite{Ras60,Byc84}, the transparency of S/F interfaces, the spin relaxation, and the disorder degree of the system. For our purposes, the exact dependence of $\varphi_0$ on these parameters is not as important as the geometry of the device. If we assume a two-dimensional SOC with momenta in the plane of the F film, and the charge current flows in $x$-direction, the phase shift $\varphi_0$ is proportional to the $y$-component of the magnetic moment according to~\cite{Buz08,Kon09,Kon15,Ber15}
\begin{equation}\label{phi0}
\varphi_0=r\frac{M_y}{M},
\end{equation}
where $M=\sqrt{M_x^2+M_y^2+M_z^2}$ is the modulus of the magnetization vector, and the parameter $r$ quantifies the SOC strength and encloses the $\alpha$-dependence. 
Equation~\eqref{phi0} establishes a direct coupling between the magnetic moment and the supercurrent, which eventually makes the current-induced inversion of the magnetization possible.

The time evolution of the magnetization can be described in terms of the Landau–Lifshitz–Gilbert (LLG) equation~\cite{Lan35,Gil04}
\begin{equation}\label{LLG}
\frac{d\textbf{M}}{d\tau}=\frac{\gamma}{M}\left ( \textbf{M}\times\frac{d\textbf{M}}{d\tau} \right )-g_r\textbf{M}\times\textbf{H}_{\text{eff}},
\end{equation}
where $g_r$ denotes the gyromagnetic ratio. The first term on the right-hand side accounts the dissipation through the phenomenological dimensionless Gilbert damping parameter $\gamma$, while the second term represents the precession around $\textbf{H}_{\text{eff}}$, which components can be calculated as~\cite{Lif90}
\begin{equation}\label{EffectiveField_FreeEnergy}
H_{\text{eff},i}=-\frac{1}{V}\frac{\partial \mathcal{F}}{\partial M_i}, \qquad\text{with}\quad i=x,y,z.
\end{equation}
Here $V$ is the volume of the F layer and $\mathcal{F}$ is the free energy of the system, which reads as follow
\begin{equation}\label{FreeEnergy}
\mathcal{F}=-E_J\varphi I_{bias}+E_s(\varphi,\varphi_0)+E_M.
\end{equation}
Here, $E_J=\Phi_0I_c/(2\pi)$ (with $\Phi_0$ being the flux quantum), $I_{bias}$ is the external current in units of $I_c$, $E_s(\varphi,\varphi_0)=E_J[1-\cos(\varphi-\varphi_0)]$, and $E_M=-\frac{\mathcal{K}V}{2}\left ( \frac{M_z}{M} \right )^2$ is the magnetic energy that depends on the anisotropy constant $\mathcal{K}$. 
In the following, we indicate the ratio between the energy scales of the system with the parameter $\varepsilon =E_J/(\mathcal{K}V)$. 

From Eq.~\eqref{EffectiveField_FreeEnergy}, we obtain the effective magnetic field
\begin{equation}\label{effectivefield}
\textbf{H}_{\text{eff}}=\frac{\mathcal{K}}{M}\left [ \varepsilon r \sin\left ( \varphi-rm_y \right )\hat{y}+m_z\hat{z} \right ],
\end{equation}
where $m_{x,y,z}=M_{x,y,z}/M$ are the normalized components of the magnetization that have to satisfy the condition $m_x^2+m_y^2+m_z^2=1$. The LLG equations can be conveniently expressed in spherical coordinates~\cite{Rom14}, so that $m_{x,y,z}$ can be written in terms of the polar and azimuthal angles $\theta$ and $\phi$ as
\begin{eqnarray}\label{Msphericalcoord}\nonumber
m_x(\tau)&=&\sin\theta(\tau)\cos\phi(\tau)\\
m_y(\tau)&=&\sin\theta(\tau)\sin\phi(\tau)\\\nonumber
m_z(\tau)&=&\cos\theta(\tau).
\end{eqnarray}
We can also define the $\theta$ and $\phi$ components of the normalized effective field as
\begin{eqnarray}\label{AngleFieldsEffective-a}
&&\widetilde{H}_{\text{eff},\theta}=\varepsilon r\sin(\varphi-rm_y)\cos\theta\sin\phi-m_z\sin\theta\qquad\\\label{AngleFieldsEffective-b}
&&\widetilde{H}_{\text{eff},\phi}=\varepsilon r\sin(\varphi-rm_y)\cos\phi.
\end{eqnarray}
Thus, by normalizing the time to the inverse of the ferromagnetic resonance frequency $\omega_F=g_r \mathcal{K}/M$, that is $t=\omega_F\tau$, the LLG equations in spherical coordinates reduce to the following two coupled equations~\cite{Rom14}
\begin{eqnarray}\label{LLGsphericalcoord_a}
\frac{\mathrm{d} \theta}{\mathrm{d}t}=&&\frac{1}{1+\gamma^2}\left ( \widetilde{H}_{\text{eff},\phi}+\gamma\; \widetilde{H}_{\text{eff},\theta}\right )\\\label{LLGsphericalcoord_b}
\sin \theta \frac{\mathrm{d} \phi}{\mathrm{d} t}=&&\frac{1}{1+\gamma^2}\left (\gamma\; \widetilde{H}_{\text{eff},\phi}- \widetilde{H}_{\text{eff},\theta} \right ).
\end{eqnarray}

The dynamics of an overdamped SFS Josephson junction can be described in terms of the resistively shunted junction (RSJ) model~\cite{Bar82,Gua15,Spa15,Gua19} generalized to include the anomalous phase shift $\varphi_0=rm_y$~\cite{Gua20}. Including also thermal fluctuations accounted by the Gaussianly distributed, delta correlated stochastic term, $I_{\text{th}}(t)$, the dynamics of the Josephson phase $\varphi(t)$ can be described by the equation~\cite{Gua20}:
\begin{equation}\label{RSJ}
\frac{d\varphi}{dt}=\omega\left [ I_{bias}(t)-\sin\left ( \varphi-rm_y \right )+I_{\text{th}}(t)\right ]+ r\frac{d m_y}{dt}.
\end{equation}
Here, the time is still normalized to the inverse of the ferromagnetic resonance frequency and $\omega=\omega_J/\omega_F$, with $\omega_J=2\pi I_c R/\Phi_0$ being the characteristic frequency~\cite{Bar82} of the junction with a normal-state resistance $R$.

The noise term $I_{\text{th}}(t)$ is a sort of ``thermal current'' with the usual white-noise statistical properties that, in normalized units, can be expressed as~\cite{Bar82,Gua13,Gua16}
\begin{eqnarray}
\left \langle I_{\text{th}}(t) \right \rangle&=&0\\
\left \langle I_{\text{th}}(t)I_{\text{th}}({t}') \right \rangle&=&2\DI \delta \left ( t-{t}' \right ),
\label{thermalcorrelatorI}
\end{eqnarray}
where we introduced the dimensionless amplitude of thermal current fluctuations defined as
\begin{equation}\label{thermalcurrentamplitude}
\DI =\frac{k_BT}{R}\frac{\omega_F}{I_c^2}=\frac{1}{\omega}\frac{k_BT}{ E_J}.
\end{equation}
In this work we assume only thermal fluctuations directly affecting the phase dynamics. Instead, the case of a stochastic ``thermal field'' $H_{\text{th}}$, with an intensity $\Dh$, included also in Eq.~\eqref{LLG}~\cite{Bro63,Cof12,Rom14,Nis15} was discussed in Ref.~\cite{Gua20}. Since the normalized intensities of the two noise sources are proportional, \emph{i.e.}, $\Dh =(\gamma\, \varepsilon \omega)\DI $~\cite{Gua20}, one could, in principle, optimize the system parameters in such a way to make the impact of the thermal field negligible with respect to the thermal current.

In our work, we assume an SFS junction biased by a rectangular current pulse, $I_{bias}$, centered at $t_c$: 
\begin{equation}\label{currentpulse}
I_{bias}(t)=\left\{\begin{matrix}
I_{\text{max}},&& t_c-\Delta t/2\leq t\leq t_c+\Delta t/2\\
0,&&\text{elsewhere}.
\end{matrix}\right.
\end{equation}
%
%
Here, $\Delta t$ is the width and $I_{\text{max}}$ is the intensity, in units of $I_c$, of the pulse, so that the condition $I_{\text{max}}<1$ means a bias current lower than the critical value, $I_c$.

\begin{figure}[t]
\centering
\includegraphics[width=1\columnwidth]{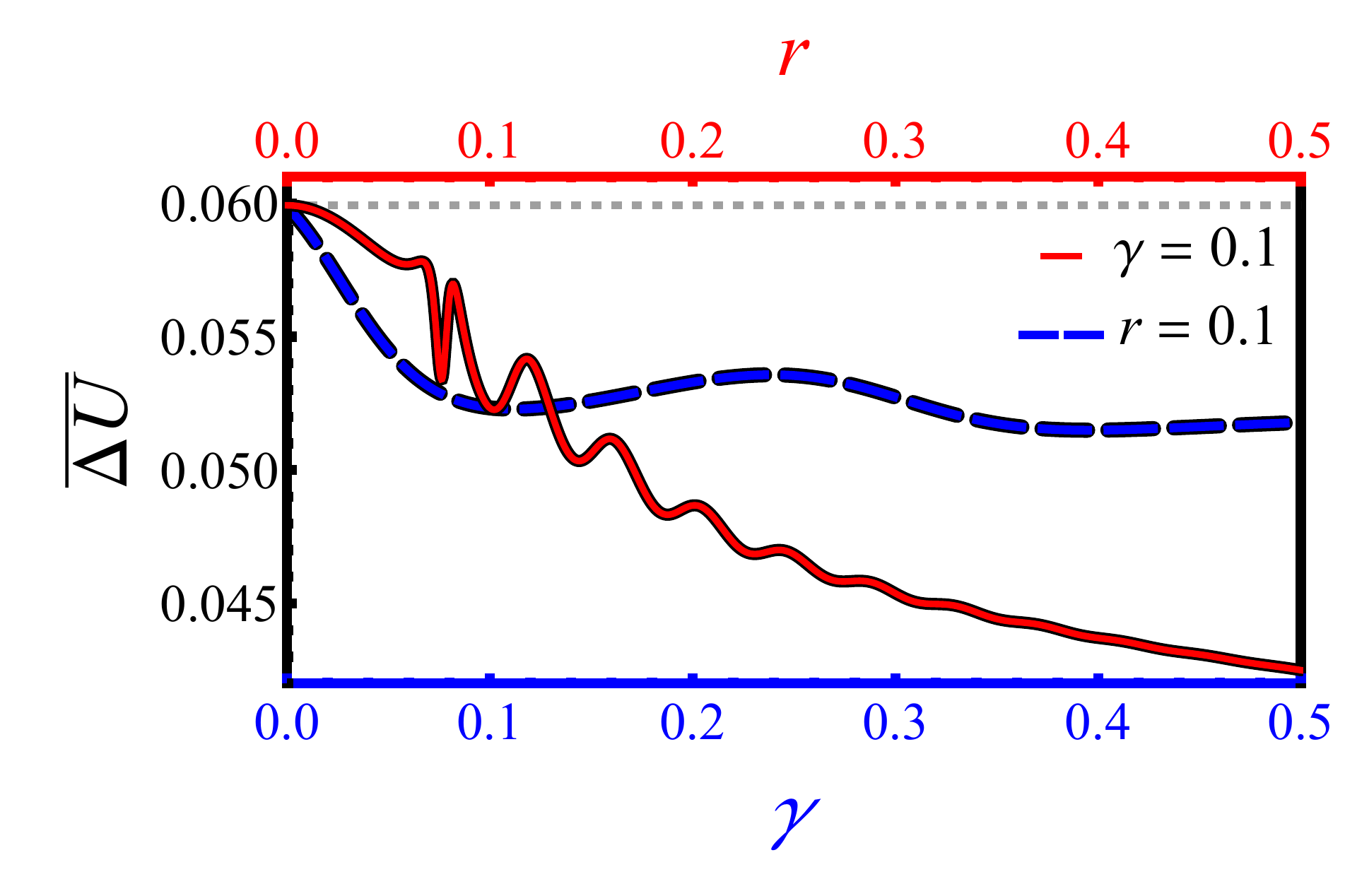}
\caption{Average washboard potential barrier, $\overline{\Delta U}(\gamma,r)$, calculated during the current pulse with intensity $I_{\text{max}}=0.9$ and duration $\Delta t=10$, versus $\gamma$ at a fixed $r=0.1$ (blue-dashed curve) and versus $r$ at a fixed $\gamma=0.1$ (red-solid curve). The horizontal gray dashed line indicates the washboard potential barrier $\Delta U(i_{b}=0.9)$. The other parameters are: $\varepsilon =10$ and $\omega=1$.} 
\label{Fig02}
\end{figure}

Often, it is useful to depict the response of a JJ in terms of the evolution of a particle, representing the superconducting phase difference $\varphi$ across the JJ, in a cosine ``washboard'' potential $U(i_b)=E_J\left [1- \cos(\varphi) -i_b\varphi\right ]$, with the current $i_b$ flowing through the junction being the slope of this potential. The resulting activation energy barrier $\Delta U(i_b)=2\left [ \sqrt{1-i_b^2} -i_b\arcsin(i_b)\right ]$ confines the phase $\varphi$ in a potential minimum. 

For completeness, we observe that the magnetization term included in the modified RSJ model affects also the height of the potential barrier. In fact, the total current contribution in Eq.~\eqref{RSJ} can be assumed to be the sum of the current pulse $I_{bias}(t)$ and the additional term stemming from the time derivative of the $y$ component of the magnetization, $I_{m_y}(t)=\frac{r}{\omega}\frac{d m_y}{d t}$. This means that during the current pulse, the phase particle ``sees'' a time-dependent potential barrier that depends on $\gamma$ and $r$. In Fig.~\ref{Fig02} we show the average potential barrier, $\overline{\Delta U}(\gamma,r)$, calculated during a current pulse with intensity $I_{\text{max}}=0.9$ and duration $\Delta t=10$, as a function of $\gamma$ and $r$. It is evident that $I_{m_y}(t)$ tends to reduce the average barrier height with respect to the value $\Delta U(I_{\text{max}})$ calculated neglecting the additional magnetic contribution, see the gray dashed line. 
This means that, by increasing $\gamma$ and $r$, the phase particle would experience an effective potential barrier slightly reduced for the enhancing of the total bias current due to the magnetic contribution. This mechanism should be taken into account when comparing the noise intensity at which the system becomes unstable with the effective potential barrier. In fact, this modulation of the washboard potential accounts the response to noisy fluctuations at low values of $\gamma$ and $r$, see Sec.~\ref{noiseeffects}.

In the following, we investigate how thermal noise affects the magnetization reversal, setting specific combinations of the system parameters. 
Specifically, we explore the response of the magnetization, with and without taking into account noise effects, by varying $\gamma$ and $r$ in suitable ranges. The energy and timescales ratios are fixed to the values $\varepsilon =10$ and $\omega=1$, respectively. 
\begin{figure}[t!!]
\centering
\includegraphics[width=\columnwidth]{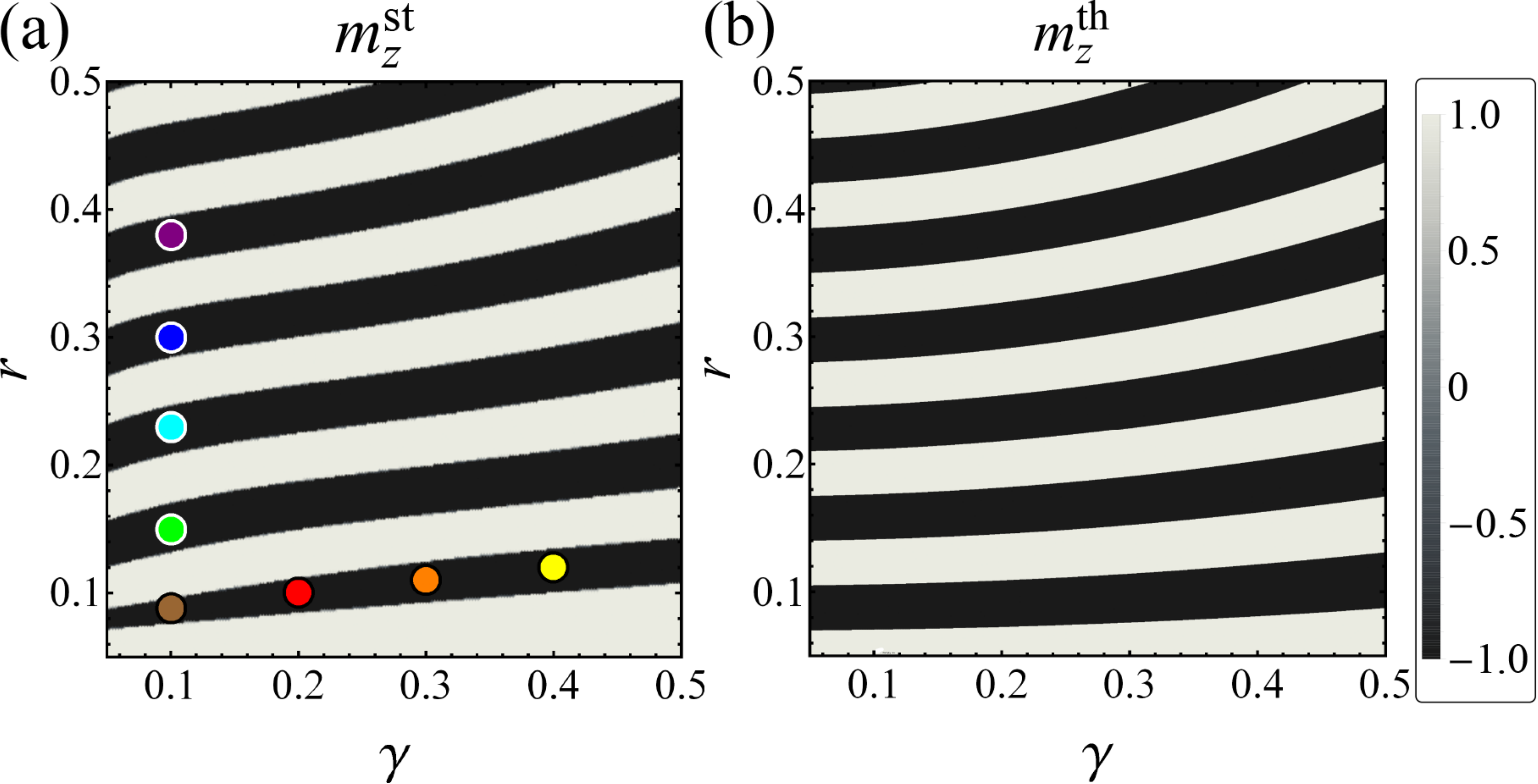}
\caption{(a) Stationary value, $m_z^{\text{st}}$, [computed by numerical solution of Eqs.~\eqref{LLGsphericalcoord_a}-\eqref{RSJ}] and (b) expected behavior, $m_z^{\text{th}}$, [calculated through Eq.~\eqref{mzth}] of the $z$-component of the magnetization as a function of $r$ and $\gamma$, in the absence of noise fluctuations, $\DI=0$. The other parameters are: $\varepsilon =10$, $\omega=1$, $I_{\text{max}}=0.9$, $\Delta t=10$, and $m_z(t=0)=+1$. The colored circles in panel (a) indicate the regions around the $(\gamma,r)$ combinations used to obtain the curves in Fig.~\ref{Fig03}. The legend in panel (b) refers to both panels.} 
\label{Fig01}
\end{figure}

\section{Results}
\label{Sec03}

\begin{figure*}[t!!]
\includegraphics[width=2\columnwidth]{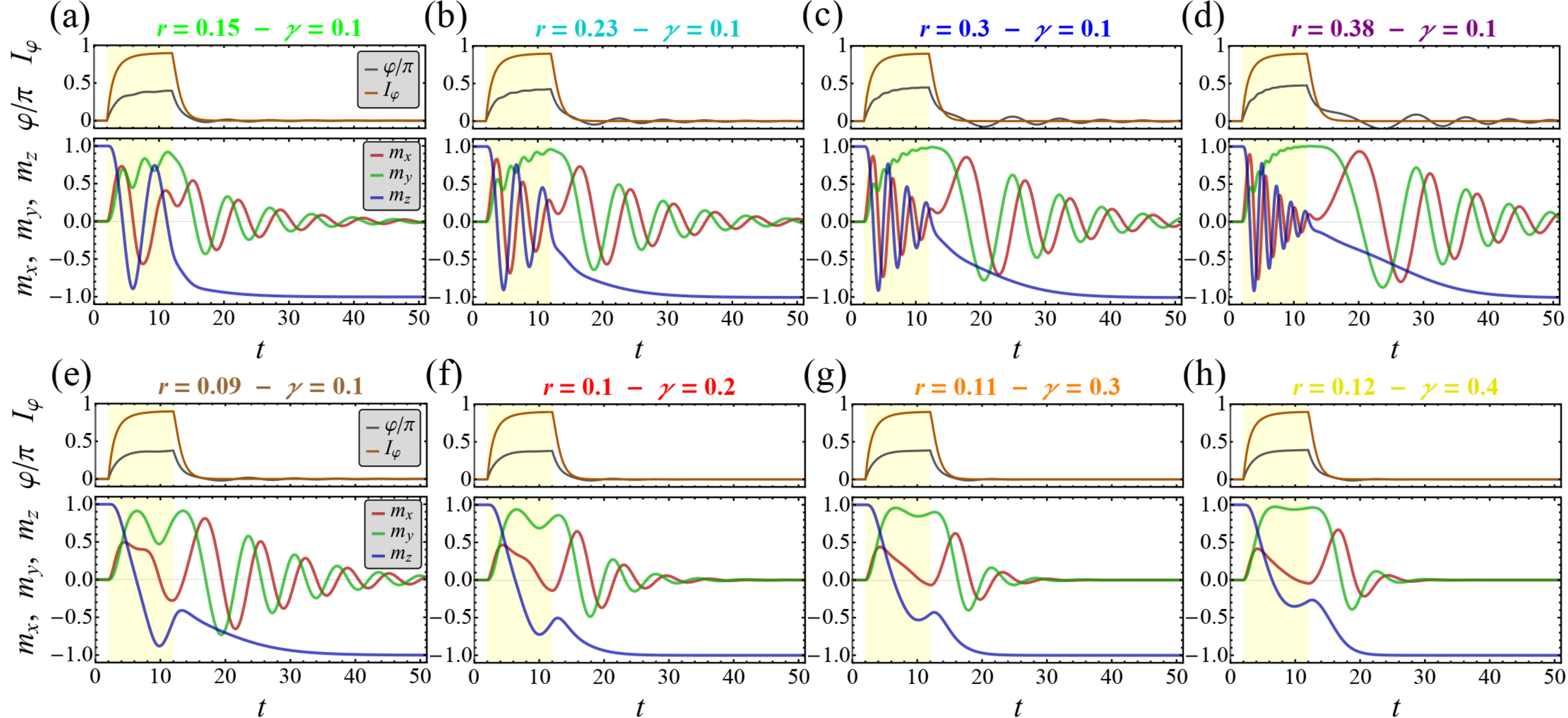}
\caption{Time evolution of Josephson phase and current, top panels, and magnetization components, bottom panels, at different $(\gamma,r)$ values, in the absence of noise, $\DI=0$. The yellow shaded regions indicate the time window in which the current pulse, with $I_{\text{max}}=0.9$ and $\Delta t=10$, is switched on. Legends in panel (a) refer to all panels.} 
\label{Fig03}
\end{figure*}

In this section, we first study the magnetic response in the absence of noise, {\it i.e.}, we impose $\DI=0$. We assume that the magnetization initially points towards the $z$-direction, that is $\textbf{M}=(0,0,1)$ at $t=0$. With this initial condition, we solve Eqs.~\eqref{LLGsphericalcoord_a}-\eqref{RSJ} self-consistently, choosing different values of the system parameters. From the solution we determine the magnetization direction after the current pulse, and in particular we focus on the stationary magnetization $m_z^{\text{st}}$, that is the value of $m_z$ long after the current pulse is turned off (at the time $t_{\text{max}}=100$).

In the following we analyze also the normalized magnetization switching time, $\tau_{\text{SW}}$, which is defined as the time (in units of $\omega_F^{-1}$) that $m_z$ takes to switch from the initial value $+1$ to $-1$, after the current is turned off. This characteristic time closely depends on the values of $r$ and $\gamma$; in particular it is possible to find specific combinations of these parameters making $\tau_{\text{SW}}$ shorter. 
This means the possibility to minimize the magnetization switching time by adjusting specific system parameters, such as $r$ and $\gamma$, by choosing materials with the proper characteristics, for instance with a larger SOC to make $r$ larger. This aspect becomes crucial designing applications based on the magnetization reversal phenomenon in a $\varphi_0$--junction, in which a high response speed can play a key role, {\it e.g.}, superconducting memory applications~\cite{GuaSol17,Gua20}.

In Fig.~\ref{Fig01}(a) we show the overall behavior of $m_z^{\text{st}}$ as a function of both the Gilbert damping parameter, $\gamma$, and the SOC strength, $r$, in response to a current pulse with intensity $I_{\text{max}}=0.9$ and width $\Delta t=10$. This contour plot is formed by a dark-bands pattern, namely, we observe regions of the $(\gamma,r)$ parametric space in which the magnetization reversal systematically occurs, {\it i.e.}, in which $m_z^{\text{st}}=-1$, and other regions in which this effect is systematically lacking, {\it i.e.}, $m_z^{\text{st}}=+1$. In other words, when increasing $r$ at a fixed $\gamma$, we observe a regular sequence of $m_z^{\text{st}}=+1$ and $m_z^{\text{st}}=-1$ values. Similar $m_z^{\text{st}}$ patterns were observed and discussed also in Refs.~\cite{Ata19,Gua20}. 
Here we go a step forward, and clarify how the magnetic response depends on the choice of $(\gamma,r)$ combinations by analyzing different regions of the density plot in Fig.~\ref{Fig01}(a). Essentially, we proceed as follows: \emph{i}) first, we choose $(\gamma,r)$ laying in different regions of the density plot (\emph{i.e.}, we fix $\gamma$ and change $r$) and \emph{ii}) then, we focus on $(\gamma,r)$ points situated on the same region (\emph{i.e.}, we fix $r$ and change $\gamma$). For more clarity, in Fig.~\ref{Fig01}(a) we also highlight with colored circles the regions around the specific $(\gamma,r)$ combinations chosen to fully characterize the temporal evolution of the magnetization and to explore noise effects.

The full time evolution of the observables of interest, {\it i.e.}, the Josephson phase, the supercurrent, and the magnetization components, in response to a current pulse with amplitude $I_{\text{max}}=0.9$ and width $\Delta t=10$ is shown in Fig.~\ref{Fig03}. 
We conveniently examine first the behavior of the phase and the Josephson current. In all panels of Fig.~\ref{Fig03}, we observe that during the current pulse, {\it i.e.}, within the yellow shaded region, the Josephson phase first increases, and then it goes to zero as the pulse is turned off. In fact, in the washboard-like picture~\cite{Bar82}, the tilting imposed by the current pulse, even considering the $I_{m_y}(t)$ contribution, is not enough to allow the ``particle'' to overcome the nearest potential barrier and to switch the system to the finite voltage ``running'' state. This means that the phase-particle remains confined within a potential minimum, so that when the current is turned off, the tilting of the washboard potential goes back to zero and the phase restores its initial position, {\it i.e.}, $\varphi\to0$. Also the Josephson current follows a similar evolution, which is characterized by a rapid exponential increase (decrease), approaching the value $I_{\text{max}}$ ($0$), when the pulse is switched on (off).

The analysis of the temporal evolution of all the components of magnetization can give us a deeper understanding of how, and why, the phenomenon of reversal of magnetization occurs. Thus, in the top panels (a)-(d) of Fig.~\ref{Fig03}, we show how the magnetization switching develops by choosing $(\gamma,r)$ combinations lying on different dark bands of the contour plot in Fig.~\ref{Fig01}(a). This means to keep fixed the Gilbert damping parameter (in particular, we set $\gamma=0.1$) and to adjust the $r$ value, {\it e.g.}, we impose $r=\{0.15,0.23,0.3,0.38\}$. 
During the current pulse, {\it i.e.}, within the yellow shaded regions, $m_x$ and $m_z$ undergo exponentially damped oscillations around a zero value. Instead, $m_y$ follows an exponential growth rippled by small, damped oscillations. In other words, the current pulse temporarily causes the precession of the magnetization around the $y$-direction. Subsequently, as the pulse is switched off, the dynamics of the magnetization moments change, indeed both $m_x$ and $m_y$ undergo damped oscillations around a zero value, while the $z$-component, after a transient regime, flips definitively to the value $m_z=-1$. We also observe that the oscillating behavior of $m_y$ is reflected in the evolution of the Josephson phase $\varphi$, which shows minima/maxima in the same position as $m_y$.

From the top panels of Fig.~\ref{Fig03}, it is evident that, as $r$ increases, also the oscillation frequency of both $m_x$ and $m_z$ increases. Furthermore, at a larger $r$, the curve of $m_y$ approaches earlier the value $1$ and its oscillations are more dampened. 
Thus, we can speculate that the dark bands in Fig.~\ref{Fig01}(a) ``differ'' essentially in the number of oscillations made by $m_z$ before reversing completely.

\begin{figure}[t!!]
\centering
\includegraphics[width=1\columnwidth]{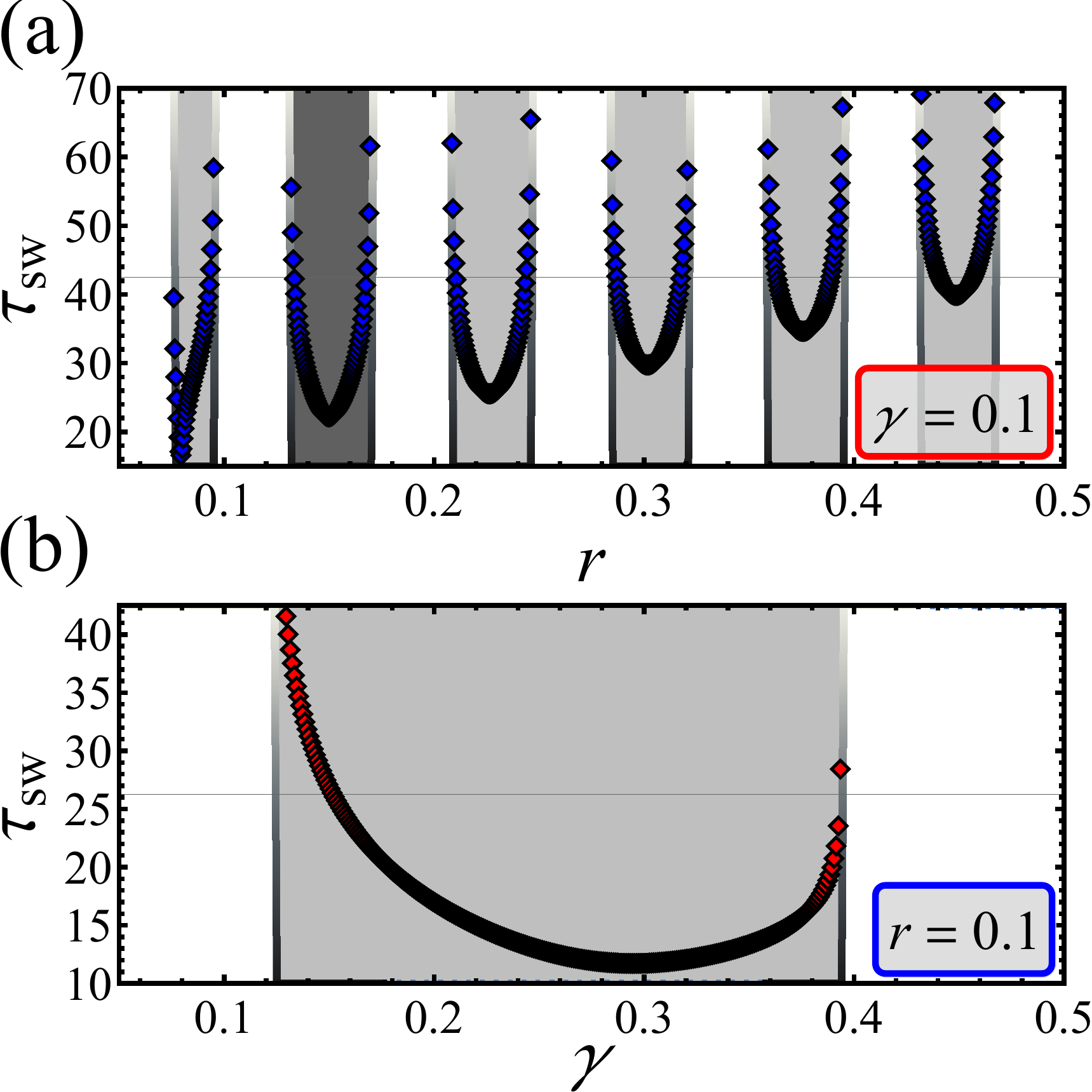}
\caption{Magnetization switching time $\tau_{\text{SW}}$, as a function of $r$ at a fixed $\gamma=0.1$, (a), and as a function of $\gamma$ at a fixed $r=0.1$, (b). In both panels, the gray shaded bands represent the range of $r$ and $\gamma$ values within which the magnetization reversal phenomenon takes place. } 
\label{Fig07}
\end{figure}
\begin{figure*}[t!!]
\centering
\includegraphics[width=2\columnwidth]{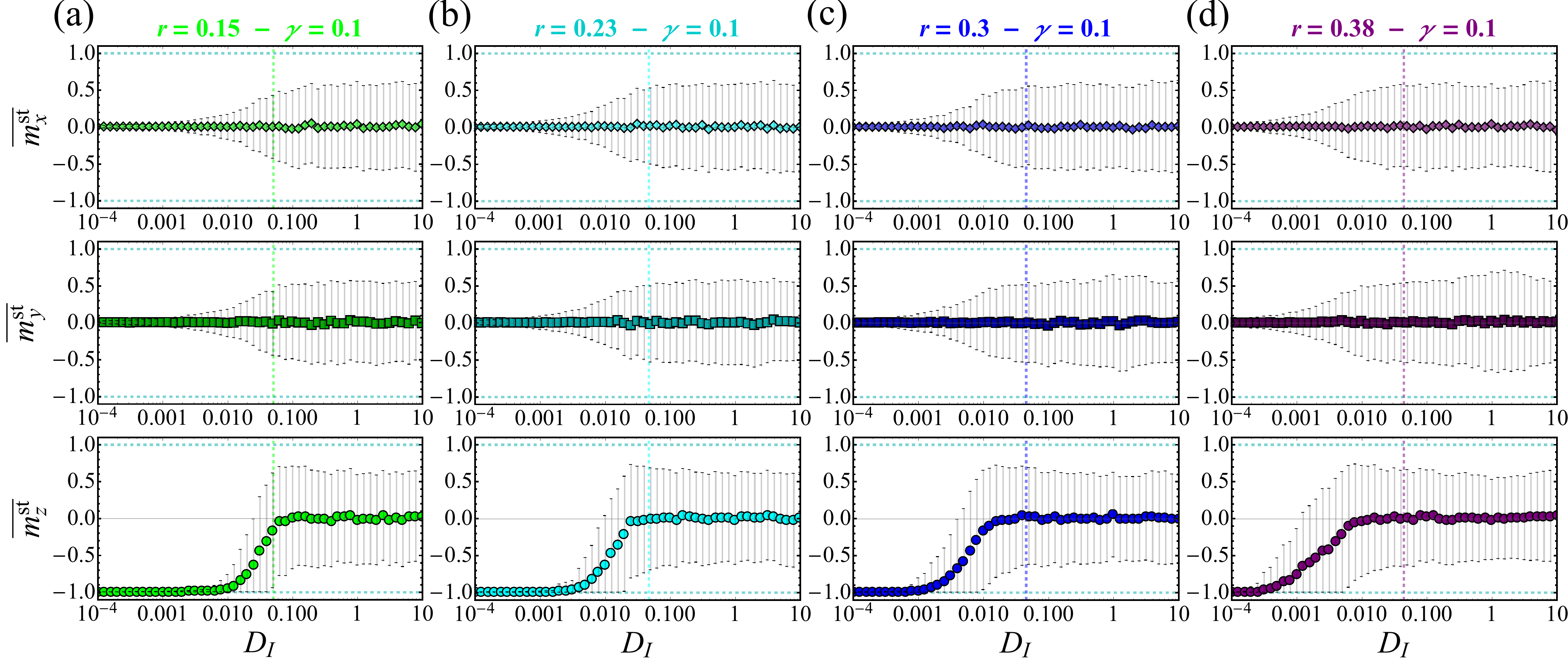}
\caption{Average stationary magnetizations, $\overline{m_x^{\text{st}}}$, $\overline{m_y^{\text{st}}}$, and $\overline{m_z^{\text{st}}}$, as a function of the thermal noise intensity, $\DI$, at the $(\gamma,r)$-values used to obtain the top panels of Fig.~\ref{Fig03}, calculated by averaging over $N_{\text{exp}}=10^3$ independent numerical repetitions. The vertical dashed lines indicate the average washboard potential barrier $\overline{\Delta U}(\gamma,r)$.} 
\label{Fig04}
\end{figure*}

Now, in order to understand how the Gilbert damping parameter affects the magnetization response, we consider $(\gamma,r)$ combinations lying on the same band of the contour plot in Fig.~\ref{Fig01}(a), see panels of (e)-(h) of Fig.~\ref{Fig03}. To this aim, we impose $r\sim0.1$ and $\gamma=\{0.1,0.2,0.3,0.4\}$. An increase of $\gamma$ tends to dampen more the oscillations of both $m_x$ and $m_y$, especially when the current is turned off. Moreover, the value of $\gamma$ affects also the time the magnetization needs to fully reverse. 
In fact, from Fig.~\ref{Fig03} one can also estimate the normalized magnetization switching time, $\tau_{\text{SW}}$.

In Fig.~\ref{Fig07}(a) and (b) we illustrate the behavior of $\tau_{\text{SW}}$ as a function of $r$ at a fixed $\gamma=0.1$ and as a function of $\gamma$ at a fixed $r=0.1$, respectively. The gray shaded fringes in this figure highlight the ranges of $r$ and $\gamma$ values within which the magnetization reversal takes place. In other words, each gray region corresponds to a dark band in Fig.~\ref{Fig01}(a). In Fig.~\ref{Fig07}(a) we can note that, within each fringe, $\tau_{\text{SW}}$ behaves non--monotonically, inasmuch it grows significantly at the edges of the gray region and reaches a local minimum roughly in its middle point. Moreover, the value of this $\tau_{\text{SW}}$ minimum tends to increase passing from a fringe to the next one at higher $r$ values. 
A non-monotonic trend emerges also looking the behavior of $\tau_{\text{SW}}$ as a function of $\gamma$, see Fig.~\ref{Fig07}(b). In particular, the $\tau_{\text{SW}}$ versus $\gamma$ curve is asymmetric and reaches the minimum value $\tau^{\text{min}}_{\text{SW}}\simeq11$ at $\gamma\simeq 0.3$. Thus, by assuming a typical ferromagnet resonance frequency equal to $\omega_F\simeq10\;\text{GHz}$~\cite{Shu19}, we obtain a minimum switching time of the order of nanoseconds.

Finally, Fig.~\ref{Fig03} suggests also that the stationary value of the magnetization depends on the state of $m_z$ when the current pulse is switched off. In fact, the magnetization reversal seems to occur when $m_z$ is in the first half of its oscillation period just when current is turned off. This simple observation allows us to roughly predict the overall response of the stationary magnetization without solving Eqs.~\eqref{LLGsphericalcoord_a}-\eqref{RSJ} numerically. In fact, according to the dynamics illustrated in Fig.~\ref{Fig03}, one could in principle assume for the $x$-- and $z$--components of the magnetization an oscillating, damped time evolution, such as $m_x(t)\simeq e^{-\frac{t}{\tau_d}}\sin(\Omega\, t)$ and $m_z(t)\simeq e^{-\frac{t}{\tau_d}}\cos(\Omega\, t)$ (where $\Omega$ is the oscillation frequency and $\tau_d$ is the exponential-decay constant). Inserting these tentative solutions in the equation for $\dot{m}_y$ that can be derived from Eq.~\eqref{LLG}, we obtain the following equation for the frequency, $\Omega(\gamma,r)$
\begin{equation}
\frac{\gamma}{1+\gamma^2}\frac{\sin(2\Omega\, t)}{2}+\Omega=I_{\text{max}}\frac{\varepsilon r}{1+\gamma^2}.
\label{omegaequation}
\end{equation}
To obtain this equation we assume that $\tau_d\gg1$ and $\sin\left ( \varphi-rm_y \right )\sim I_{\text{max}}$. These two assumptions correspond, essentially, to suppose a small $r$ and a pulse duration longer than the exponential rise of the Josephson current, see Fig.~\ref{Fig03}.
Finally, by comparing the oscillation period, $T_{_{\Omega}}(\gamma,r)=2\pi/\Omega(\gamma,r)$, and the pulse duration $\Delta t$, we can give an estimate of the stationary magnetization as following
\begin{equation}
m_z^{\text{th}}(\gamma,r)\!=\!\left\{\begin{matrix}
-1\quad\text{if}\quad \text{Mod}\left [ \Delta t,T_{_{\Omega}}(\gamma,r) \right ]\leq T_{_{\Omega}}(\gamma,r)/2\;\,\\ 
+1\quad\text{if}\quad \text{Mod}\left [ \Delta t,T_{_{\Omega}}(\gamma,r) \right ]> T_{_{\Omega}}(\gamma,r)/2,
\end{matrix}\right.
\label{mzth}
\end{equation}
where $\text{Mod}[a,b]$ gives the remainder on division of $a$ by $b$. The behavior of $m_z^{\text{th}}$ as a function of $\gamma$ and $r$ is shown in Fig.~\ref{Fig01}(b), at $I_{\text{max}}=0.9$, $\Delta t=10$, and $\varepsilon=10$. This contour plot recalls closely that obtained solving Eqs.~\eqref{LLGsphericalcoord_a}-\eqref{RSJ} numerically and shown in Fig.~\ref{Fig01}(a), especially at high $(\gamma,r)$ values. This means that the simple assumptions made to obtain Fig.~\ref{Fig01}(b) allow to grasp, in our case, the essential features behind the magnetization reversal phenomenon. As a closing remark, we observe that an analytical solution for the magnetization dynamics induced by a current pulse was recently proposed in Ref.~\cite{Maz20}. In this work, Mazanik \emph{et al.} formulate criteria for magnetization reversal in a $\varphi_0$ junctions, obtaining a good agreement between numerical results and analytical prediction, in specific ranges of the system parameters.
\begin{figure}[b!!]
\centering
\includegraphics[width=\columnwidth]{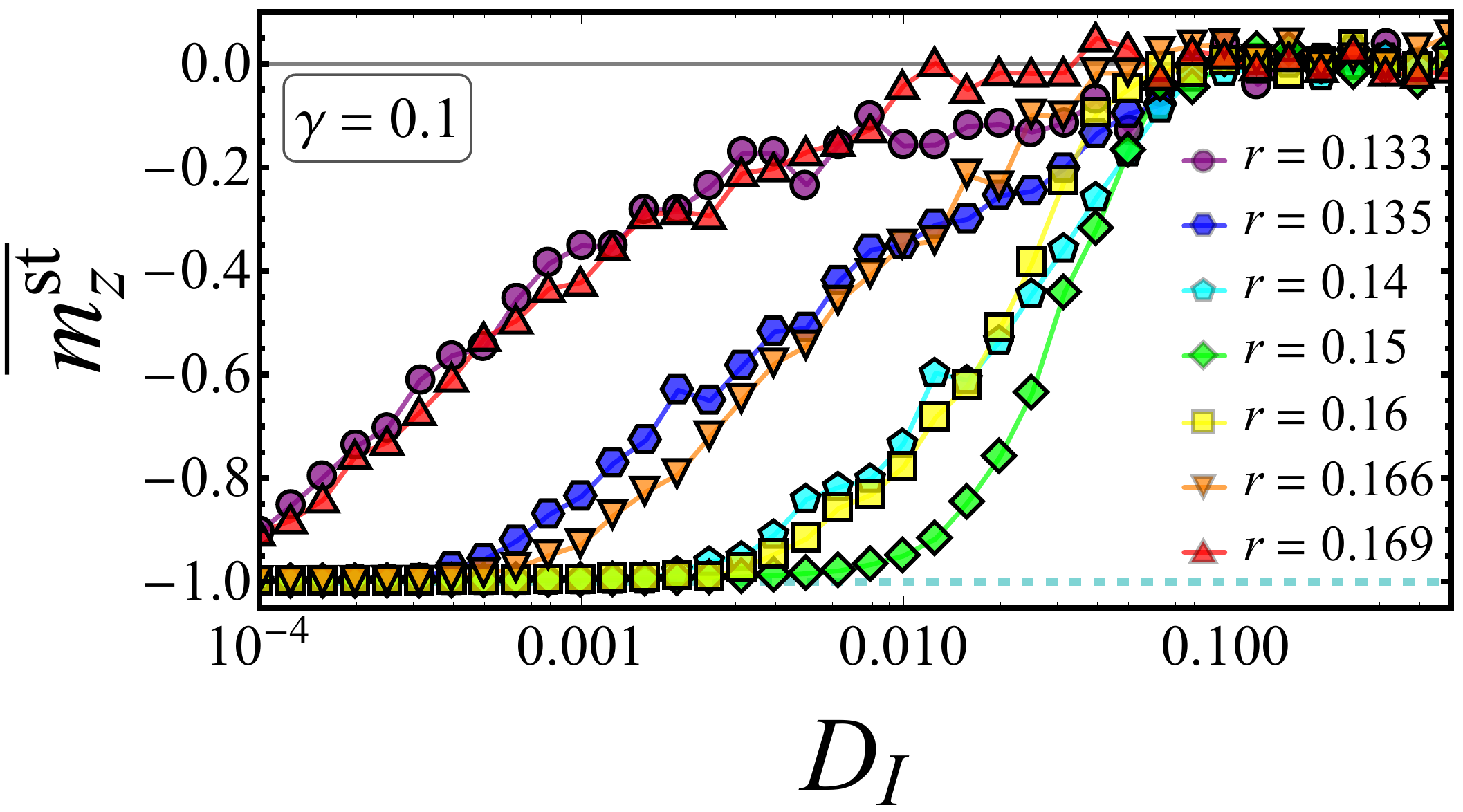}
\caption{Average stationary magnetization, $\overline{m_z^{\text{st}}}$, as a function of the thermal noise intensity, $\DI$, at $\gamma=0.1$ and $r\in(0.133-0.17)$, calculated by averaging over $N_{\text{exp}}=10^3$ independent numerical repetitions. Lines in the figure are guides for the eye.} 
\label{Fig08}
\end{figure}

\subsection{Noise effects}
\label{noiseeffects}

In this section we discuss the effects produced by a non-negligible thermal noise source on the magnetization dynamics. The temperature can influence significantly the response of the system, since thermal fluctuations may eventually induce an unwanted magnetization switch or prevent the magnetization reverse. We investigate the effects of stochastic thermal fluctuations in the phase dynamics, by including a Gaussian noise source in the RSJ model, see Eq.~\eqref{RSJ}.

In the following, we focus on the average components of the stationary magnetization, $\overline{m_x^{\text{st}}}$, $\overline{m_y^{\text{st}}}$, and $\overline{m_z^{\text{st}}}$, which are computed by averaging over $N_{\text{exp}}=10^{3}$ independent numerical repetitions in the presence of a non-negligible thermal noise intensity, $\DI\neq0$. 

Figure~\ref{Fig04} shows the behavior of the average magnetization as a function of the noise intensity $\DI$, obtained by fixing $\gamma=0.1$ and varying $r$; in particular we choose the $(\gamma,r)$ combinations used to obtain the curves in the top panels of Fig.~\ref{Fig03}. In this way we explore the noise effects by focusing on the different gray bands in Figs.~\ref{Fig01} and~\ref{Fig07} at a given $\gamma$.
We indicate with $D^0_{\text{I}}$ the noise amplitude at which $\overline{m_z^{\text{st}}}$ starts to deviate significantly from the value $-1$ and with $\overline{D}_{\text{I}}$ the noise amplitude at which the switching process is fully suppressed, that is when $\overline{m_z^{\text{st}}}$ approaches the zero value. 
In all panels of Fig.~\ref{Fig04} we also mark with a dashed vertical line the $\DI$ value coinciding with the average barrier height, $\overline{\Delta U}(\gamma,r)$. 

From Fig.~\ref{Fig04}(a), which is obtained for $r=0.15$ and $\gamma=0.1$, one sees that $\overline{m_z^{\text{st}}}\simeq-1$ only for $\DI\lesssim D^0_{\text{I}}=0.01$. For higher noise intensities both $\overline{m_z^{\text{st}}}$ and the error bars increase, approaching a zero value of $\overline{m_z^{\text{st}}}$ exactly for $\overline{D}_{\text{I}}\simeq\overline{\Delta U}(\gamma,r)=0.051$. 
By increasing further the noise intensity, \emph{i.e.}, for $\DI\gtrsim\overline{D}_{\text{I}}$, $\overline{m_z^{\text{st}}}$ still remains close to zero, showing however quite large error bars that indicate a highly fluctuating response totally driven by noise. 
The values of $\overline{m_x^{\text{st}}}$ and $\overline{m_y^{\text{st}}}$ hover around zero at each noise intensity, despite their error bars tend to enlarge when $\DI\gtrsim\overline{D}_{\text{I}}$. 

Increasing $r$, both the value of $D^0_{\text{I}}$ and $\overline{D}_{\text{I}}$ reduces significantly, see Fig.~\ref{Fig04}(b-d), so that the greater $r$, the more $\overline{D}_{\text{I}}$ deviates from $\overline{\Delta U}$. This means that, increasing $r$, the system is more sensitive to noise and the maximum temperature at which it can reside, without significant effects on the stationary magnetization, reduces. In other words, the robustness against thermal fluctuations of the magnetization reversal effect is damaged by a high $r$ value.

\begin{figure*}[t!!]
\centering
\includegraphics[width=2\columnwidth]{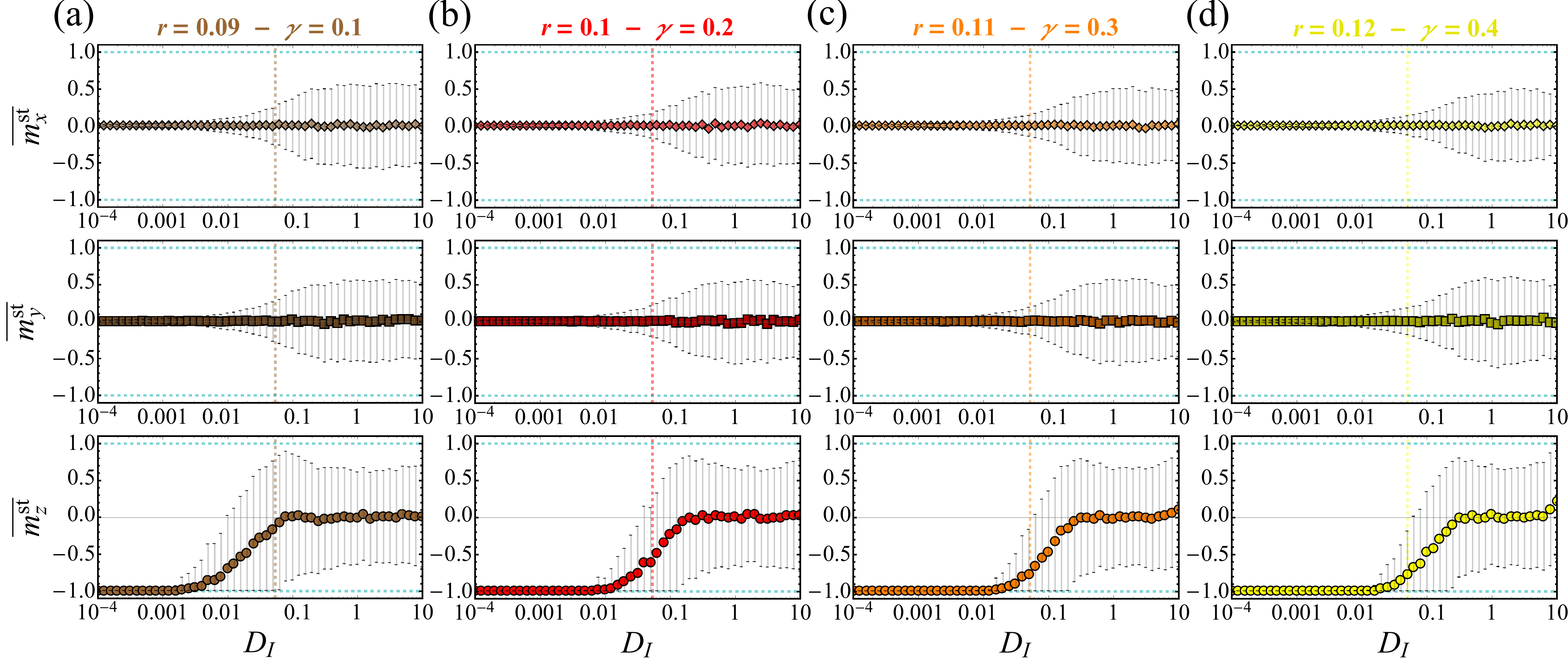}
\caption{Average stationary magnetizations, $\overline{m_x^{\text{st}}}$, $\overline{m_y^{\text{st}}}$, and $\overline{m_z^{\text{st}}}$, as a function of the thermal noise intensity, $\DI$, at the $(\gamma,r)$-values used to obtain the bottom panels of Fig.~\ref{Fig03}, calculated by averaging over $N_{\text{exp}}=10^3$ independent numerical repetitions. The vertical dashed lines indicate the average washboard potential barrier $\overline{\Delta U}(\gamma,r)$.} 
\label{Fig06}
\end{figure*}

To obtain the curves in Fig.~\ref{Fig04}, we fixed $\gamma$ while $r$ changes in such a way to explore the system response in the different gray bands of Fig.~\ref{Fig07}(a), more specifically, we chose the $r$ values lying exactly in the midpoint of each band. On the other hand, we can demonstrate that, if we restrict now to just a single gray band, the impact of thermal noise can change significantly even for a small variation of $r$. For instance, in Fig.~\ref{Fig08} we show how $\overline{m_z^{\text{st}}}$ versus $\DI$ modifies by setting $r\in(0.13,0.17)$, with $\gamma=0.1$, that is we focus on the darkest fringe in Fig.~\ref{Fig07}(a). We observe that imposing $r=0.15$ noise effects are kept at a minimum, while thermal fluctuations affect more the magnetization switching if $r$ is slightly increased, or decreased. 
In particular, the value of $\overline{D}_{\text{I}}$ depends little on $r$, unlike the value of $D^0_{\text{I}}$ which changes significantly by changing $r$. Specifically, for $r=\{0.15$, $0.14$, and $0.135\}$ we obtain the values $D^0_{\text{I}}\simeq\{10$, $3$, and $0.5\}\times10^{-3}$, respectively. 
If we suppose a temperature--dependent critical current, with $I_c=10\;\mu\text{A}$ at low temperatures~\footnote{In the case of weak proximity effect and large exchange field in F, the critical current temperature-dependence is proportional to $I_c(T)\propto\Delta(T)\tanh \left [ \Delta(T)/(k_BT) \right ]$~\cite{Ber05}, where $\Delta (T)$ is the superconducting gap. Thus, to find the temperature corresponding to a given noise intensity, we use this relation, with a zero-temperature value $I_c(0)=10\;\mu\text{A}$.}, these noise intensities $D^0_{\text{I}}$ correspond to the normalized temperatures $T/T_c\simeq\{0.85$, $0.58$, and $0.12\}$, respectively.
In summary, at a fixed $\gamma$, the optimal value of $r$ corresponds to the midpoint of a gray band, whereas just a small change of $r$ is enough to undermine the stability of the system.

Finally, we discuss how an increase of $\gamma$ can influence the magnetization reversal. In Fig.~\ref{Fig06} we present the average stationary magnetizations $\overline{m_x^{\text{st}}}$, $\overline{m_y^{\text{st}}}$, and $\overline{m_z^{\text{st}}}$ versus $\DI$, imposing the $(\gamma,r)$ combinations used to obtain the curves in the bottom panels of Fig.~\ref{Fig03}. For $\gamma=0.1$, $\overline{m_z^{\text{st}}}$ approaches a zero value only for $\overline{D}_{\text{I}}\simeq\overline{\Delta U}(\gamma,r)=0.054$, see Fig.~\ref{Fig06}(a) . For higher noise intensities, $\overline{m_z^{\text{st}}}$ remains close to zero, showing quite large error bars. The same happens for the $x$ and $y-$components of the magnetization. 
The $\gamma$ coefficient acts as a friction on the magnetization dynamics, so that a larger $\gamma$ means a system more ``stiff'' and, therefore, less sensitive to noisy disturbances. This is why the noise intensity $\overline{D}_{\text{I}}$, at which $\overline{m_z^{\text{st}}}$ approaches a zero value, increases with $\gamma$, see Figs.~\ref{Fig04}(b)-(d). In particular, at high $\gamma$ values, we observe that $\overline{D}_{\text{I}}$ is always well above $\overline{\Delta U}(\gamma,r)$. This means that, in view of a possible application based on a $\varphi_0$--junction, a larger $\gamma$ could allow a higher working temperature, still preserving the magnetization reversal phenomenon. In principle, one could think to optimize the system parameters in such a way to make, for instance, the switching time minimal, see Fig.~\ref{Fig07}, still keeping the device at a suitable working temperature.

\section{Conclusions}
\label{Sec04}

In conclusion, in this paper we discuss the bistable magnetic response of a current-biased $\varphi_0$--junction, that is a superconductor -- ferromagnet -- superconductor Josephson junction with a Rashba-like spin-orbit coupling. The direction of the magnetization of the ferromagnetic layer can be inverted via controlled current pulses. We study the temporal evolution of all the components of the magnetization in different conditions. We determine the values of intrinsic system parameters, such as the Gilbert damping and strength of the spin-orbit coupling, corresponding to a minimum switching time of the magnetization. We also suggest a way to grasp readily the essential features behind the magnetization reversal phenomenon through simple assumptions, without facing the numerical solutions of the differential equations for both the magnetization and the Josephson dynamics. 

We also explore how the magnetization switching time strongly depends on the $(\gamma,r)$ values. We observe a switching times of order nanoseconds, which can be acceptable for qubits~\cite{Rev11}, but could be too slow for conventional low temperature electronics. Also the specific shape of the driving pulse and the orientation of the switching field can significantly affect the magnetization switching time~\cite{Pan08}.
However, in this work, we aim to demonstrate that the switching time can strongly depend on the value of the parameter $(\gamma,r)$. From the other side, we assumed fixed energy and timescales ratios. Usually, the energy ratio $\varepsilon$ ranges from $\varepsilon\sim100$~\cite{Kon09}, if the magnetic anisotropy is weak, to $\varepsilon\sim1$~\cite{Shu19}, in the case of a stronger anisotropy. In our calculation we choose an intermediate value, $\varepsilon=10$. The typical ferromagnet resonance frequency is $\omega_F\sim 10\;\text{GHz}$, while the characteristic Josephson frequency is usually in the range $\omega_J\in[10-100]\;\text{GHz}$. In our work, we conservatively choose $\omega=\omega_J/\omega_F = 1$, but we could reasonably impose also a higher value for the frequency ratio $\omega$. In this case, one can expect a strong suppression of the magnetization switching time (e.g., see Ref.~\cite{Maz20}. Please note that in this paper the frequency ratio is defined as $w=\omega_F/\omega_J$, so that $w=\omega^{-1}$). Since $\omega_J=2\pi/\Phi_0 I_c R$, the magnetization switching time could be reduced by adjusting the values of the junction parameters $I_c$ and $R$. Moreover, in this work we estimate a magnetization switching time of the order of nanoseconds by assuming a ferromagnet resonance frequency equal to $\omega_F\sim 10\;\text{GHz}$; since $\omega_F\propto \mathcal{K}/M$ (with $\mathcal{K}$ and $M$ being the anisotropy constant and the modulus of the magnetization vector, respectively) one could envisage to reduce further this ratio to obtain a shorter magnetization switching time.

Finally, we explore the robustness of the current-induced magnetization reversal against thermal fluctuations, in order to find the regime of system parameters in which the magnetization switching induced by a current pulse is more stable. In particular, we demonstrate that the choice of a low $r$ and/or a high $\gamma$ value can be convenient to keep thermal fluctuations at bay, in order to increase the temperature at which the system can reside still preserving the magnetization reversal effect. 

The investigation of thermal effects on hybrid superconductive/ferromagnetic structures is important in applications of novel electronic devices, such as spintronics~\cite{Lin15,Gin16}, qubits and superconducting logic elements~\cite{Feo10}, and detectors with electron cooling~\cite{Kuz19}. Moreover, the role of noise can become crucial in high-speed switching electronics, where stabilization effects due to noise can lead to enhancement of the switching time, the so-called noise-delayed switching effect. Indeed, noise-enhanced stabilization effects can play a relevant role by reducing the size of the magnetic element~\cite{Smi10} and are also demonstrated to be important in the switching dynamics of Josephson devices~\cite{Val14,GuaVal15,GuaVal16}.

\section{Acknowledgments}
\label{Sec05}
F.S.B. acknowledge funding by the Spanish Ministerio de Ciencia, Innovaci\'on y Universidades (MICINN) (Project No. FIS2017-82804-P), partial support by Grupos Consolidados UPV/EHU del Gobierno Vasco (Grant No. IT1249-19), and by EU's Horizon 2020 research and innovation program under Grant Agreement No. 800923 (SUPERTED).

\printcredits




\end{document}